# How really effective are Multimodal Hints in enhancing Visual Target Spotting? Some evidence from a usability study.

Suzanne Kieffer[1] and Noëlle Carbonell[2].

[1]BCHI – Université Catholique de Louvain – Place des Doyens, 1 – B1348 Louvain-La-Neuve, Belgium

Phone: +32(0)10/478379, Fax: +32(0)10/47.83.24, Email: kieffer@isys.ucl.ac.be

[2]LORIA - Campus Scientifique - BP 239 - F54506 Vandoeuvre-lès-Nancy Cedex, France

Phone: +33(0)383592032, Fax: +33(0)383413079, Email: Noelle.Carbonell@loria.fr

## Abstract.

The main aim of the work presented here is to contribute to computer science advances in the multimodal usability area, in-as-much as it addresses one of the major issues relating to the generation of effective oral system messages: how to design messages which effectively help users to locate specific graphical objects in information visualisations? An experimental study was carried out to determine whether oral messages including coarse information on the locations of graphical objects on the current display may facilitate target detection tasks sufficiently for making it worth while to integrate such messages in GUIs. The display spatial layout varied in order to test the influence of visual presentation structure on the contribution of these messages to facilitating visual search on crowded displays. Finally, three levels of task difficulty were defined, based mainly on the target visual complexity and the number of distractors in the scene. The findings suggest that spatial information messages improve participants' visual search performances significantly; they are more appropriate to radial structures than to matrix, random and elleptic structures; and, they are particularly useful for performing difficult visual search tasks.

### Categories and Subject Descriptors

H.5.2 [User Interfaces]: Ergonomics, Evaluation/Methodology, Graphical User Interfaces (GUI), Natural Language, Voice I/O.

I.3.6 [Methodology and Techniques]: Interaction Techniques.

*Keywords: Visual search. Multimodal system messages. Speech and graphics. Usability study. Experimental evaluation. Visual target spotting.*

# Context and motivation

In the 90s, numerous interactive information visualisation techniques were invented and commercialized, mainly in order to facilitate user access to graphical information and multimedia. Numerous visualisation techniques have been propounded, such as, zoom views, multiscale interfaces and hierarchical views (see, [6] for a general overview of visualisation techniques and their use). Few ergonomic studies have been published about the effectiveness of visual search in such spatial organisations.

At the same time multimodal interaction appeared. Numerous forms of speech-based input multimodality have been proposed, implemented and tested. Combinations of speech with gestural modalities have been studied extensively, especially combinations of speech with modalities exploiting new input media, such as touch screens, pens, data gloves, haptic devices. Both usability and implementation issues have been considered; see, among others, [4] on speech and pen for the first category of issues, [3] for the second category. Contrastingly, speech combined with text and graphics has only motivated a few studies. As an output modality, speech is mostly used either as a substitute for standard visual presentation modes (cf. phone services) or for supplementing deficiencies in visual exchange channels.

In spite of that, most information visualisations now combine both images and spoken language. One essential question remains. Do verbal messages provide a specific and significant contribution to human-computer interaction for visual search on complex displays? Information visualisations convey more indications to users than other media, even when combined [7]. Consequently, it is not obvious that the combination of spoken information and visual information will improve and facilitate visual activites within information visualisations. Moreover, it is justifiable to ask whether such output multimodality will not result in a cognitive overload.

# Current study objectives

The main aim of the work presented here is to contribute to computer science advances in the multimodal usability area, in-as-much as it addresses one of the major issues relating to the generation of effective oral system messages: how to design messages which effectively help users to locate specific graphical objects in information visualisations?

An experimental study was carried out to determine whether oral messages including coarse information on the locations of graphical objects on the current display may facilitate target detection tasks sufficiently for making it worth while to integrate such messages in GUIs. In addition, the display spatial layout varied in order to test the influence of visual presentation structure on the contribution of these messages to facilitating visual search on crowded displays. Finally, three levels of task difficulty were defined, based mainly on the target visual complexity and the number of distractors in the scene [1].

Target detection was selected as the experimental task for the following reasons. First, it is one of the few human activites, besides reading, that have motivated a significant amount of psychological research, cf. [1]. Second, the design of numerous computer applications may benefit from a better knowledge of this activity such as: online help for current interactive application software, geographical applications and navigation systems in vehicles, visualisations of very large data sets, navigation in hierarchical views of personal data sets.

## Methodology

Twenty-four volunteers participated to this study. This gender-balanced group of participants was composed of experienced computer users, ranged in age from 24 to 29 years and with normal eyesight (assessed using the Bioptor test kit). Thus, all participants were expert mouse users with alike quick motor reactions and they were experienced in visual search activities. Target selection time and spotting accuracy were likely to reflect visual search performance reliably, and task learning effects were prevented.

### Procedure

First, participants were given an explanation of the research study. Prior to the experiment, they performed the Bioptor eye test, filled in demographic questionnaire and performed some training target selection tasks. Finally, participants performed the series of target mouse selections. Following the computer-based tasks, participants were asked to fill in second questionnaire and were debriefed.

### Experimental Design

The usability study employed a 4x3x10 factorial design, with 2 modality conditions: 120 scenes were used in each modality condition, each scene including 30 photographs, organized along one out of four standard symmetrical structures (see figure 1). Three levels of task complexity were used (easy, difficult, and very difficult). Levels of task difficulty were distributed among the four structures. In short, the usability study employed 10 scenes by level of difficulty (3) and structure (4) in each modality condition. Participants were asked to retrieve and select, with the mouse, as fast as they could, a pre-viewed photograph in each scene according to 2 modality conditions: the VP condition (target visual presentation) and the MP condition (target multimodal presentation). In the VP condition, the isolated target was displayed in the centre of the screen during 3 seconds. In the MP condition, a short oral message containing information on the target location was played simultaneously with the target visual presentation.

Messages were composed of one or two short spatial phrases, for instance, "On the left (of the screen)" or "At the bottom (of the screen), on the right". Following target presentation, participants had to click on a button in the centre of the screen for launching the scene display. Thus, the position of the mouse at the beginning of the search was identical for all tasks.

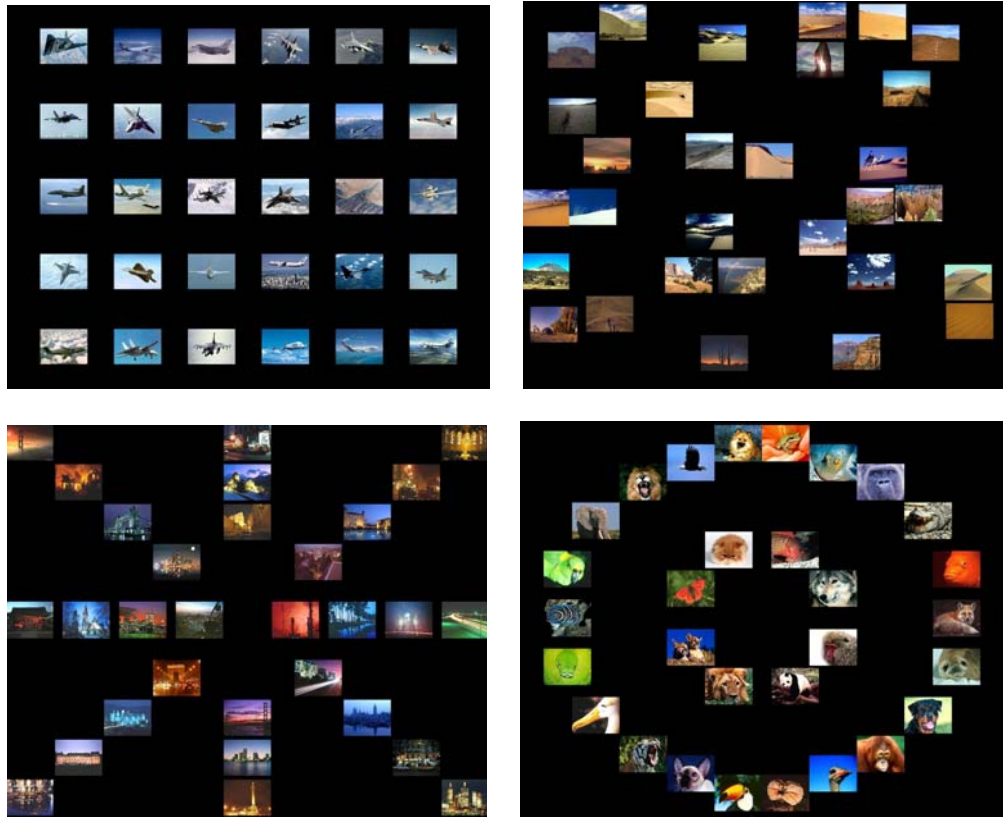

Figure 1: Matrix, Random, Elliptic and Radial structures.

The order of modality conditions was counterbalanced between participants so as to neutralize possible task learning effects.

The dependent variables used to assess participant task performance were target selection time (from scene display onset until first mouse click), and accuracy (i.e., mouse click on the target vs elsewhere).

## Results

### Analyses Methodology

Analyses of variance (ANOVA) were used to examine the presence of significant differences in task performance, as measured by both selection times and error numbers, according to: (i) the target presentation mode; (ii) the scene structure; (iii) the task difficulty. If significant differences were revealed by the ANOVA procedure, then paired t-tests were also performed to further highlight any difference. The following analyses were computed over all subjects (24): selection times and errors per presentation mode; selection times per presentation mode and structure; selection times per presentation mode and task difficulty.

The results of the ANOVA procedure are presented in tables 1 and 2. Tables 3, 4 and 5 present a summary of participants' performance analyses respectively per presentation mode, per presentation mode and structure, and, per presentation mode and task difficulty. In tables 4 and 5, values preceded by "-" or "+" are respectively inferior or superior to the corresponding average values per presentation mode reported in table 2.

**Target presentation mode and task difficulty are highly significant factors**

The results from table 1 show that target presentation mode, scene structure and task difficulty are significant factors. First, they validate our classification of scenes into three levels of difficulty (40 scenes per target presentation mode and level). Secondly, they suggest that target presentation mode and task difficulty have more influence on results than scene structure. In order to elicit the specific influence of scene structure on participants' selection times with and without spatial information messages -respectively, in the MP condition vs. in the VP condition- a complementary ANOVA procedure was computed. Results are reported in table 2 together with results concerning task difficulty.

Table 1. ANOVA Procedure. Factors: target presentation mode, scene structure, task difficulty.

| Factors | Selection times | Error numbers |
|---|---|---|
| Presentation (i) | t=1202.98; p<.0001 | t=23.18; p<.0001 |
| Structure (ii) | t=6.26; p=0.0003 | t=2.58; p=0.05 |
| Difficulty (iii) | t=32.49; p<.0001 | t=7.59; p=0.0005 |

The results from table 2 reveal that spatial information messages suppress scene structure influence on participants' speed (MP: t=2.2602; p=0.0899). On the other hand, task difficulty still has an effect on participants' speed with spatial information messages as well as without any message (VP: t=22.72, p<.0001; MP: t=21.50, p<.0001).

Table 2. Compementary ANOVA Procedure. Factors: scene structure, task difficulty.

| Factors | Presentation | Selection times |
|---|---|---|
| Structure | VP | t=4.89; p=0.0022 |
| | MP | t=2.2602; p=0.0899 |
| Difficulty | VP | t=22.72; p<.0001 |

| | MP | t=21.50; p<.0001 |
|---|----|------------------|

**Multimodal assistance impoves target detection significantly**

Spatial information messages improve participants' visual search performances significantly (table 3). Averaged target selection times computed over all participants are thrice shorter in the MP condition than in the VP condition (1747 ms versus 5674 ms). This result is highly significant (t=-34.07; p<.0001). In addition, one observes twice less errors in the MP condition than in the VP condition (79 versus 150). This result is highly significant (t=23.656; p<.0001).

Moreover, participants expressed very positive judgments on multimodal target presentations, both in the questionnaires and during the debriefing interviews. For 75% of them (18), target spotting was easier (less hesitations) in the MP condition than in the VP condition. Most participants mentioned that they had experienced some strain and visual fatigue during the VP condition whereas they had felt perfectly comfortable during the MP condition. All participants considered that oral messages including coarse information on target location could provide efficient support to visual search activities, and two thirds (16) expressed a marked preference for the MP condition.

Table 3. Participants' selection times and errors per target presentation mode.

| Presentation mode | Avg ST ms | Std Dev ms | Nb Errors | % Errors | Nb Obs |
|-------------------|-----------|------------|-----------|----------|--------|
| VP | 5674 | 5985 | 150 | 5.2% | 2880 |
| MP | 1747 | 1552 | 79 | 2.7% | 2880 |

**Spatial information messages match radial structures best**

Radial structure improves participants' visual search performances -speed and accuracy- in the MP condition (table 4). Since scene structure is not a significant factor in the MP condition, no paired t-test was computed. However, compared to matrix, random and elliptic structures, the radial structure leads to the fastest and most accurate target detection. This result may be explained by the following reason: spatial information messages like "On the left (of the screen)" or "At the bottom (of the screen), on the right" are more appropriate to radial structures than matrix, elliptic and random structures. This interpretation is supported by some spontaneous comments collected during the debriefing interviews: the radial structure was preferred by 11 out of 24 participants, that is to say almost half of them.

In the VP condition, selection time differences between the radial and elliptic structures, the radial and matrix structures, the elliptic and matrix structures are statistically significant (respectively, t=3.64 p=0.0003, t=1.25 p=0.0024, t=2.18 p=0.0296). The four spatial structures can be ordered as follows according to increasing averaged selection times: radial, random, matrix, elliptic. These results are somewhat unexpected, since participants were experienced computer users, and the use of 2D arrays is currently prevailing for displaying pictures. Participants' subjective judgments were at variance with their performances: more than half expressed a marked preference for elliptic layouts compared to the other structures, and two thirds of them judged either the matrix or the radial structure the most inefficient layout. Participants' performances and subjective judgments concerning the matrix structure in the VP condition are in accordance with the results presented in [6].

Table 4. Participants' selection times and errors per structure.

| Structure | Presentation | Avg ST ms | Std Dev ms |
|-----------|--------------|-----------|------------|
| Radial    | VP           | -5081     | -5565      |
|           | MP           | -1640     | -1256      |
| Random    | VP           | -5626     | -5819      |
|           | MP           | -1737     | -1437      |
| Matrix    | VP           | +5738     | -5879      |
|           | MP           | +1763     | +1819      |
| Elliptic  | VP           | +6250     | +6585      |
|           | MP           | +1851     | +1633      |

**Influence of messages depends on the task complexity**

Spatial information messages are particularly useful for performing difficult visual search tasks. Acually, a careful analysis of participants' performances shows that average selection times increase from level 1 (easy) to level 3 (very difficult) less rapidly in the MP condition (25%) than in the VP condition (35%) (See table 5). Therefore, it seems worth while to assist users in difficult visual search activities through spatial information messages. As such short oral messages will be well accepted by potential users, or so it seems according to participants' subjective judgments, their use for helping users to carry out easy visual search tasks may also be considered.

In both conditions, averaged selection times increase noticeably from level 1 to level 3. For the VP condition, the difference between any pair of levels is statistically significant, the difference between levels 1 and 3 being highly significant (t=-6.40; p<.0001). For the MP condition, differences between level 1 and 3, and 2 and 3, are highly significant (t=-5.29; p<.0001 and t=-5.33; p<.0001 respectively), while the difference between levels 1 and 2 do not reach significance.

Table 5. Participants' selection times and errors per target presentation mode and task difficulty.

| Difficulty | Presentation | Avg ST ms | Std Dev ms | Actual Errors | % Errors |
|---|---|---|---|---|---|
| Easy | VP | -4919 | -5011 | 17 | 18% |
| | MP | -1611 | -1387 | 10 | 20% |
| Difficult | VP | -5439 | -5879 | 33 | 34% |
| | MP | -1620 | -1272 | 15 | 31% |
| Very difficult | VP | +6663 | +6801 | 46 | 48% |
| | MP | +2012 | +1893 | 24 | 49% |

## Conclusion and discussion

The usability study reported here aims at assessing the actual contribution of voice system messages to visual serach efficiency and comfort. The experimental task used was target detection which is representative of visual search tasks since it is commonly used in current GUI environments. Oral messages comprised one or two spatial phrases conveying coarse information on the target location on the display. Participants carried out visual search tasks in two conditions differing from each other in initial target presentation only: visual presentation of the target versus multimodal presentation, that is, visual presentation of the target simultaneously with oral indications on its location on the screen.

Spatial information messages improve participants' visual search performances significantly. In addition, regarding participants' speed and accuracy, they match radial structures best compared to matrix, random and elleptic structures. Finally, spatial information messages are particularly useful for performing difficult visual search tasks. According to subjective judgments, oral messages were well accepted, and multimodal target presentations were preferred to visual presentations by a majority of participants.

Consequently, designers of graphical user interfaces might consider resorting to short oral messages including coarse spatial information for drawing users' attention to some displayed object. As such messages are likely to be well accepted by users, they may provide designers of advanced conversational user interfaces with a useful substitute "pointing" technique for any visual enhancement method in interaction contexts where gaze activity is intense and there is a risk of visual attention overload and eyestrain.

However, these empirical results need to be consolidated and further refined before reliable recommendations inferred from them can be proposed to designers. Indeed, the target detection task proposed to participants during the experiment remains a laboratory task, even though it is realistic. Our further research needs to focus on more elaborate tasks than locate and identify, such as compare, associate, distinguish, rank, cluster, correlate or categorize. In particular, a case study exhibiting effective and beneficial integration of multimodal messages to a domain-oriented interactive system would be extremely convincing for potential adopters who are left wondering what performance improvement would be achieved with the multimodal interface. These multimodal usability challenges are comparable with those expressed within the Information Visualization community [5].

## Prospective applications of multimodal messages

The main difficulty for applying our spatial information multimodal messages to target detection within present GUIs lies in that the system is supposed to know exactly where the potential targets are located. Nevertheless, the following case study would offer the opportunity to successfully test the technical faisability of this new form of human-computer interaction: combination of speech and visual presentation as an output style of multimodality. It deals with the valuation of the potential benefits from speech, as a graphical expression mode in a context of image retrieval [2]. In their work, Descampe et al. use a coarse-to-fine classification process in order to retrieve similar textures in mega-images (JPEG 2000), that is, textures are similar to a sample entered by the user. Spatial information multimodal messages could be used to improve the whole retrieval process by helping the user to retrieve collections of similartextures within mega-images. The long-term view of this application is to take advantage from HCI and Signal Processing knowledge to design, implement and evaluate a multimodal user interface.